\documentclass[11pt, a4paper]{article}
\pdfoutput=1

\usepackage[usenames, dvipsnames]{color}
\usepackage[T1]{fontenc}
\usepackage[latin1]{inputenc}
\usepackage{lmodern}
\usepackage{amsmath}
\usepackage[pdftex]{graphicx}
\usepackage{lastpage}
\usepackage{amssymb} 
\usepackage{hyperref}
\usepackage{cite}
\usepackage{soul}
\usepackage{mathtools}

\usepackage{subcaption}

\usepackage{esvect}
\renewcommand*\vec{\vv}

\usepackage{xspace}
\newcommand*{\ie}{i.e.\@\xspace}
\newcommand*{\eg}{e.g.\@\xspace}

\newcommand*{\Eq}{Eq.\@\xspace}

\newcommand*{\Wlog}{W.l.o.g.\@\xspace}

\newcommand{\ex}{\text{e}}
   
\usepackage{slashed}
\usepackage{braket}
\usepackage{simplewick}
\usepackage{bbm}
\newcommand\idop{\mathbbm 1}

\usepackage{mathtools}
\usepackage{leftidx,tensor}

\usepackage{xifthen}
\newcommand*\diff{\mathrm{d}} 
\newcommand*\ldiff[2][]{ \ifthenelse{\isempty{#1}}{ \diff #2}{\diff^#1#2} \,} 
\let\limitint\int 
\renewcommand{\int}{\limitint \!} 

\usepackage{subcaption}

\usepackage{titling}
\usepackage{authblk}

\title{The Scales of the Infrared}

\author[1,2]{C\'{e}sar G\'{o}mez\thanks{cesar.gomez@uam.es}}
\author[1]{Raoul Letschka\thanks{raoul.letschka@csic.es}}
\author[2,3]{Sebastian Zell\thanks{sebastian.zell@campus.lmu.de}}
\affil[1]{Instituto de F\'{i}sica Te\'orica UAM-CSIC, Universidad Aut\'onoma de Madrid, \mbox{Cantoblanco, 28049 Madrid, Spain}}
\affil[2]{Arnold Sommerfeld Center, Ludwig-Maximilians-Universit\"at, \mbox{Theresienstra\ss e 37, 80333 M\"unchen, Germany}}
\affil[3]{Max-Planck-Institut f\"ur Physik, F\"ohringer Ring 6, 80805 M\"unchen, Germany}

\newcommand*\B{B_{\alpha,\, \beta}} 
\newcommand*\Bpp{B_{\alpha',\, \beta'}}
\newcommand*\Bb{B_{\beta,\, \beta'}}
\newcommand*\Ba{B_{\alpha,\, \alpha'}}
\newcommand*\Bp{B_{\alpha,\, \beta'}}
\newcommand*\Bpa{B_{\alpha',\, \beta}}  

\renewcommand*\S[1][]{ \ifthenelse{\isempty{#1}}{S_{\alpha,\, \beta}}{S_{\alpha,\, \beta #1}}}
\newcommand*\SD[1][]{ \ifthenelse{\isempty{#1}}{\tilde{S}_{\alpha,\, \beta}}{\tilde{S}_{\alpha,\, \beta #1}}}
\newcommand*\Sp[1][]{ \ifthenelse{\isempty{#1}}{S_{\alpha,\, \beta'}}{S_{\alpha,\, \beta'#1}}}
\newcommand*\Spp[1][]{ \ifthenelse{\isempty{#1}}{S{\alpha',\, \beta'}}{S_{\alpha,\, \beta'#1}}}
\newcommand*\G{\Gamma_{\alpha,\, \beta}}

\newcommand\braD[1]{\prescript{r}{\lambda}{\langle\bra{#1}}}
\newcommand\ketD[1]{\ket{#1}\rangle_\lambda^r}

\newcommand*{\Dad}{\ket{D(\alpha)}_\lambda^r}
\newcommand*{\Da}{\prescript{r}{\lambda}{\bra{D(\alpha)}}}
\newcommand*{\Dbd}{\ket{D(\beta)}_\lambda^r}
\newcommand*{\Dabd}{\ket{\gamma(\alpha, \beta)}_r^\epsilon}

\newcommand*{\Dabp}{\prescript{\epsilon}{r}{\bra{\gamma(\alpha, \beta')}}}

\newcommand*\F[1]{\mathcal{F}^{(l)}_{\alpha,\, \beta}(#1)}
\newcommand*\Fp[1]{\mathcal{F}^{(l)}_{\alpha,\, \beta'}(#1)}
\newcommand*\Fb[1]{\mathcal{F}^{(l)}_{\beta,\, \beta'}(#1)}
\newcommand*\FCa[1]{\mathcal{F}^{(l)}_{\alpha}(#1)}
\newcommand*\FCb[1]{\mathcal{F}^{(l)}_{\beta}(#1)}

\newcommand*{\W}[1]{\hat{W}(#1)}

\let\oldsqrt\sqrt
\def\sqrt{\mathpalette\DHLhksqrt}
\def\DHLhksqrt#1#2{%
\setbox0=\hbox{$#1\oldsqrt{#2\,}$}\dimen0=\ht0
\advance\dimen0-0.2\ht0
\setbox2=\hbox{\vrule height\ht0 depth -\dimen0}%
{\box0\lower0.4pt\box2}}

\begin{document}

\allowdisplaybreaks

\maketitle

\vspace{\baselineskip}
\begin{abstract}
		 In theories with long-range forces like QED or perturbative gravity, loop corrections lead to vanishing amplitudes. There are two well-known procedures to address these infrared divergences: dressing of asymptotic states and inclusion of soft emission.  Although both yield the same IR-finite rates, we point out that they are not equivalent since they encode different infrared scales.  In particular, dressing states are independent of the resolution scale of radiation. Instead, they define radiative vacua in the von Neumann space. After a review of these concepts, the goal of this paper is to present a combined formalism that can simultaneously describe both dressing and radiation. This unified approach allows us to tackle the problem of quantum decoherence due to tracing over unresolved radiation.  We obtain an IR-finite density matrix with non-vanishing off-diagonal elements and estimate how its purity depends on scattering kinematics and the resolution scale. Along the way, we comment on collinear divergences as well as the connection of large gauge transformations and dressing.
		
\end{abstract}

\newpage

\tableofcontents

\section{Introduction}
\subsubsection*{Combination of Inclusive and Dressed Formalism}
In its original presentation \cite{weinberg64}, soft theorems simply fix the structure of amplitudes when the momentum $\vec{k}$ of an asymptotic massless boson of spin $1$ or $2$ vanishes. In the case of a photon, we obtain for instance
\begin{equation} \label{softTheorem}
\lim_{|\vec{k}|\rightarrow 0} \S[\vec{k}]^{(l)} = \frac{\F{\vec{k}}}{|\vec{k}|^{1/2}}\S \,,
\end{equation}
where $S_{\alpha,\beta}$ is the amplitude without soft emission and $l$ is the polarization of the soft emitted boson. The function $\F{\vec{k}}$ arises from the sum of the different possible ways in which the soft mode can be emitted from the external lines. It can be straightforwardly computed by Taylor expanding the propagators of  nearly on-shell charged particles. The key point of the theorem lies in the observation that the amplitude \eqref{softTheorem}, which is divergent in the $\vec{k}=0$ limit, does not satisfy Lorentz invariance unless the sum of incoming charges is equal to the sum of the outgoing ones. In other words, the soft theorem identifies what conservation law is needed in order to have a well-defined and Lorentz invariant soft limit of the amplitudes. In the case of $j=2$, \ie gravity, the conservation law is the equivalence between inertial and gravitational mass. This conclusion can be reached independently of the potential infrared divergences of $S_{\alpha,\beta}$ due to quantum loops in which the massless mode runs.

However, the full significance of soft theorems only becomes apparent after taking into account loop corrections. Their soft part factorizes and yields 
\begin{equation} \label{1loop}
	\S^{(\text{1 loop})} = \S \left(1-\frac{\B}{2} \ln \frac{\Lambda}{\lambda} \right)  \,,
\end{equation}
where $\lambda$ is an IR-regulator that will go to zero and $\Lambda$ is an inessential UV-cutoff. Moreover, $\B$ is a non-negative kinematical factor, which is zero only for trivial forward scattering. Consequently, the amplitude \eqref{1loop} is divergent in the limit $\lambda \rightarrow 0$ for all non-trivial processes. As a next step, one can consider an arbitrary number of soft loops, \ie calculate their contribution to all orders in the coupling constant. The result of this resummation is that all non-trivial amplitudes vanish. Of course, one should not conclude from this that trivial scattering processes, in which the charges of incoming and outgoing particles match at each angle, are the only ones that take place in Nature. Instead, this merely implies that there is no interaction if asymptotic states do not contain soft photons. Once we drop this unphysical restriction, we uncover non-trivial scattering processes.

In a first approach \cite{BN, YFS, weinberg}, one leaves the initial state unchanged but enlarges the final state by soft radiation, which is defined as any photon state with a total energy below some resolution scale $\epsilon$. The first order of the corresponding rate follows directly from the soft photon theorem \eqref{softTheorem}: $\Gamma^{(\text{1 emission})} =|\S|^2 \Big(1+ \limitint_{\lambda}^{\epsilon} \ldiff[3]{\vec{k}}\left|\F{\vec{k}}\right|^2/|\vec{k}| \Big)$.
The important result of infrared physics is that the integral of $\F{\vec{k}}$ yields the same factor $\B$ as the loop computation, \ie the rate for the emission of one photon is
\begin{equation} \label{1emission}
\Gamma^{(\text{1 emission})} = \left|\S\right|^2 \left(1+  \B\ln \frac{\epsilon}{\lambda} \right) \,.
\end{equation}
Therefore, the sum of the corrections due to loops and due to emission is finite. As for loops, one can also resum the contribution of emission to all orders and it turns out that the same cancellation of divergences persists. Therefore, taking into account soft emission leads to a finite total rate.\footnote
{The situation is fully analogous in QED and gravity. This only changes for massless electrons. Whereas gravity is insensitive to the electron mass, additional collinear divergences arise in massless QED \cite{kln}, which we shall briefly review in section \ref{ssec:comments}.}
This approach, which can be called {\it inclusive formalism}, predicts how the rate of the processes depends on the resolution scale. Moreover, it has a nice physical interpretation: Since any accelerated charge emits soft bremsstrahlung, the probability for a non-trivial scattering process without any emission is zero.

In a second approach \cite{chung,kibble,FK}, one starts from the observation that in gapless theories, asymptotic states are not eigenstates of the free Hamiltonian, but the leading order of the interaction term has to be taken into account. Therefore, charged particles in asymptotic states should be dressed by an infinite amount of soft photons. In this approach, which can be called {\it dressed formalism}, both initial and final states are analogously dressed, but the dressing is independent of interaction, \ie the dressing of the initial state only depends on the initial state and the dressing of the final state only depends on the final state. As in the inclusive formalism, one needs to introduce a new energy scale to define the dressing states, which we shall call $r$.  Since its physical meaning is not immediately apparent, we will momentarily postpone its discussion. In the dressed formalism, one can use the soft theorems to obtain  -- up to subleading corrections -- the same IR-finite rate as in the inclusive formalism, provided one sets $r = \epsilon$. As we will discuss, however, the reason for identifying these two scales is unclear.

Because of their infinite photon number, dressing states do not belong to the Fock space but can only be defined in the von Neumann space $\mathcal{H}_{\text{VN}}$ \cite{VN}. It consists of infinitely many subspaces, $\mathcal{H}_{\text{VN}}= \otimes_{\alpha}[\alpha]$,  where each subspace $[\alpha]$, dubbed equivalence class, is isomorphic to the Fock space with an inequivalent representation of the creation and annihilation operator algebra \cite{WS}. So asymptotic dynamics defines a dressing operator $\hat{W}$ that associates to a state of hard charged quanta, which we shall call $\ket{\alpha}$, a photon state $\ket{D(\alpha)}$ in the von Neumann equivalence class $[\alpha]$:
 \begin{equation} \label{dressingIntro}
\hat{W}: \ket{\alpha} \rightarrow \ket{\alpha}\rangle := \ket{\alpha} \otimes \ket{D(\alpha)} \,.
 \end{equation}
As notation indicates, the equivalence class $[\alpha]$ is sensitive to the state of the charged particle $\ket{\alpha}$, \ie the photon state $\ket{D(\alpha')}$ of a different charged particle $\ket{\alpha'}$ belongs to a different -- and orthogonal -- equivalence class $[\alpha']$. It is crucial to note that for the representation of the creation annihilation algebra in $[\alpha]$, the dressing state $\ket{D(\alpha)}$ is the vacuum. This reflects that fact that there is no radiation in the dressed formalism.
 
Since both the inclusive and the dressed formalism yield the same rate, the question arises if they are equivalent. This would come as a big surprise since the requirements of the two formalisms -- emission of bremsstrahlung versus well-defined asymptotic states -- are very different. Both requirements are, however, very reasonable and should be fulfilled. Therefore, we shall argue that both dressing and soft radiation should be present in a generic process. Thus, the first goal of this note is to present a concrete formalism that interpolates between the inclusive and the dressed formalism and makes the distinction between radiation and dressing explicit. We shall call it {\it combined formalism} and will derive it from first principles by applying the $S$-matrix, as operator in $\mathcal{H}_{\text{VN}}$, to the dressed initial state $\ket{\alpha}\rangle$. This gives
\begin{equation}\label{SMatrixIntro}
\hat{S} \ket{\alpha}\rangle = \sum_{\beta} \sum_{\gamma\in[\beta]} S_{\alpha,\beta \gamma} \ket{\beta}\rangle\otimes \ket{\gamma(\alpha, \beta)_{ D(\beta)}} \,,
\end{equation}
where $\beta$ sums over all possible final charged states. In turn, each of those determines an equivalence class $[\beta]$. The crucial novelty as compared to the dressed formalism is that a radiation state $\ket{\gamma(\alpha, \beta)_{ D(\beta)}}$, which depends on both $\ket{\alpha}$ and $\ket{\beta}$, exists on top of the  radiative vacuum defined by the dressing state $\ket{D(\beta)}$. Not surprisingly, it will turn out that also in the combined formalism, one obtains the same IR-finite rate as in the two known formalisms. 

\subsubsection*{IR-Finite Density Matrix}
This finding immediately raises the question about the relevance of our construction. However, one can go one step further than the rate and investigate the density matrix of the final state. Obviously, its diagonal is determined by the known IR-finite rates. So the task consists in determining the IR-limit of the off-diagonal  pieces of the density matrix. These elements contain the information about the quantum coherence of the final state.  For a particular simplified setup, the density matrix in the presence of soft bremsstrahlung was already studied some time ago. In a framework of real time evolution \cite{decoherence1, decoherence2}, which goes beyond the $S$-matrix description, the result was that tracing over unresolvable soft radiation leads to some loss of coherence. But for realistic timescales, the decoherence is generically small. As it should be, it consequently does not spoil the interference properties that we observe in Nature.
	
However, it was also derived in \cite{decoherence1, decoherence2} how coherence depends on the timescale $t_{\text{obs}}$, after which the final state is observed: Albeit slowly, it decreases as the timescale increases.
In the limit of infinite time, one obtains full decoherence. Since this is precisely the limit on which the definition of the $S$-matrix is based, it is immediately evident that it might be difficult to derive the density matrix from the $S$-matrix. In the inclusive formalism, this expectation turns out to be fulfilled. Tracing over soft radiation, which is required for IR-finiteness, leads to full decoherence \cite{decoherence2}. In an independent line of research, this finding has recently received renewed interest in the context of a generic scattering process \cite{semenoff, semenoff2, coherence1,semenoff3}. However, if it were not possible to improve this result, this would mean that the $S$-matrix is {\it in principle} unable to describe any interference phenomena in QED. While we have proposed a heuristic method to obtain IR-finite off-diagonal elements \cite{coherence1}, this finding is a clear indication that the inclusive formalism is insufficient to describe the density matrix of the final scattering state.

In the dressed formalism, the opposite situation is realized. The reason is that dressing photons are part of the definition of the asymptotic states and are independent of the scattering process. Therefore, there is no reason to trace over them. In fact, it is not even clear how to define the trace in the von Neumann space since it would amount to squeezing the infinite von Neumann subspaces into a single Fock space. This means that there is no tracing and no decoherence in the dressed formalism.\footnote
{In \cite{semenoff2, semenoff3}, the scales of radiation and dressing were identified, $r=\epsilon$, and a tracing over dressing states was performed. Since states in different equivalence classes are orthogonal, a similar result as in the inclusive formalism, \ie a fully decohered density matrix of the final state, was obtained. As explained, however, the physical meaning of tracing over dressing states is unclear to us.}
 Also this finding is unsatisfactory since one expects some decoherence due to the emission of unresolvable soft bremsstrahlung. 

The situation improves in the combined formalism that we propose. In it, the final state consists both of dressing, defined by the scale $r$, and of soft radiation, defined by the scale $\epsilon$. In order to obtain the density matrix of the final state, we have to trace over radiation but not over dressing. In this way,  we avoid full decoherence. Since the purity of the density matrix depends on the scale $r$, the connection to \cite{decoherence1, decoherence2} makes its meaning evident: It is set by the timescale after which the final state is observed, $r=t_{\text{obs}}^{-1}$. Thus, we obtain a sensible IR-finite density matrix, thereby continuing our work \cite{coherence1}. This is a clear indication of the physical relevance of the combined formalism. The second goal of this note therefore is to compute the density matrix of the final scattering state in the combined formalism and to estimate the amount of decoherence it exhibits. 

In section \ref{sec:theory}, we will introduce the theoretical tools upon which our analysis is based. In particular, we will review well-known results on the von Neumann space and on the definition of asymptotic states.  In doing so, our goal is not to be mathematically rigorous, but to put well-known mathematical results in a physical context with the aim of making the distinction between dressing and radiation evident. In section \ref{sec:amplitude}, we will combine computations of the inclusive and dressed formalism to determine the final state of scattering in the combined formalism. \Eq \eqref{finalStateCoherent} constitutes our result. Moreover, we will make additional comments about the inclusive formalism, collinear divergences as well as the connection of large gauge transformations and dressing. Then we will proceed in section \ref{sec:densityMatrix} to calculate the density matrix, which is displayed in \Eq \eqref{coherentDensitySuperposition}, and give a bound on its decoherence. We conclude in section \ref{sec:conclusion} and appendix \ref{app:identitiy} contains part of the calculation of the final state in our combined formalism.

\section{The Distinction of Dressing and Radiation}
\label{sec:theory}
\subsection{Introduction to von Neumann Spaces}
\label{ssec:introductionVN}
We begin by giving a brief review of how the Fock space can be constructed and what complications arise in a gapless theory. Our starting point are the Hilbert spaces in each momentum mode $\vec{k}$. So we are given well-defined Hilbert spaces $\mathcal{H}_{\vec{k}}$, which feature inner products $\braket{\ ,\ }_{\vec{k}}$ and creation and annihilation operators $\hat{a}_{l,\vec{k}}^\dagger$, $\hat{a}_{l,\vec{k}}$ that fulfill canonical commutation relations:
\begin{equation}
\left[\hat{a}_{l,\vec{k}}, \hat{a}_{l',\vec{k}}^\dagger\right] \sim \delta_{l l'} \,, \qquad \left[\hat{a}_{l,\vec{k}}, \hat{a}_{l',\vec{k}}\right] = \left[\hat{a}_{l,\vec{k}}^\dagger, \hat{a}_{l',\vec{k}}^\dagger\right] =0 \,.
\end{equation}
We already included the polarization $l$ since we  will later be interested in photons.
The problem lies in defining the tensor product $\bigotimes_{\vec{k}}\mathcal{H}_{\vec{k}}$  of the infinitely many Hilbert spaces corresponding to all possible momenta $\vec{k}$.

For this task we can rely on the seminal work by von Neumann \cite{VN}, who defined the space $\mathcal{H}_{\text{VN}} \subset \bigotimes_{\vec{k}}\mathcal{H}_{\vec{k}}$. It consists of elements for which a scalar product can be defined. For $\ket{\varphi}, \ket{\Psi} \in \mathcal{H}_{\text{VN}}$, \ie $\ket{\varphi} = \otimes_{\vec{k}}\ket{\varphi}_{\vec{k}}$ and $\ket{\Psi} = \otimes_{\vec{k}}\ket{\Psi}_{\vec{k}}$, it is given as
\begin{equation}
\braket{\varphi|\Psi} := \prod_{\vec{k}} \braket{\varphi_{\vec{k}}|\Psi_{\vec{k}}}_{\vec{k}}\,.
\end{equation}
 It is clear from this definition that the von Neumann space is very big. In particular, it contains any product of states that are normalizable in the individual $\mathcal{H}_{\vec{k}}$, \ie $\ket{\varphi} = \otimes_{\vec{k}}\ket{\varphi}_{\vec{k}}$ such that $\braket{\varphi_{\vec{k}}|\varphi_{\vec{k}}}_{\vec{k}} = 1$ for all $\vec{k}$. \Wlog we will assume normalized states from now on.

 This scalar product defines an equivalence relation in the von Neumann space given by
\begin{equation}
	\ket{\varphi} \sim \ket{\Psi} \ :\Leftrightarrow\ \sum_{\vec{k}} \left|\braket{\varphi_{\vec{k}}|\Psi_{\vec{k}}}_{\vec{k}}-1\right| \ \text{convergent.}
\end{equation}
The significance of this equivalence relation lies in the fact that elements from different equivalence classes are orthogonal,
\begin{equation}
\ket{\varphi} \nsim \ket{\Psi} \ \Rightarrow \braket{\varphi|\Psi} = 0 \,.
\end{equation}
 Therefore, the equivalence classes constitute mutually disjoint subspaces in the von Neumann space.
   The physical implications of this construction were derived in \cite{WS}.
First of all, a special role is played by the equivalence class of $\ket{0} := \otimes_{\vec{k}}\ket{0}_{\vec{k}}$, which we denote by $[0]$. In it, one has the standard representation of canonical commutation relations:
\begin{equation} \label{commutationRelationsFock}
\left[\hat{a}_{l,\vec{k}}, \hat{a}_{l',\vec{k'}}^\dagger\right] = \delta^{(3)}(\vec{k}-\vec{k}') \delta_{l l'} \,, \qquad \left[\hat{a}_{l,\vec{k}}, \hat{a}_{l',\vec{k'}}\right] = \left[\hat{a}_{l,\vec{k}}^\dagger, \hat{a}_{l',\vec{k'}}^\dagger\right] =0 \,.
\end{equation}
Then we can define the particle number operator as 
\begin{equation}
	\hat{N}:= \sum_{\vec{k},l} \hat{a}_{\vec{k},l}^\dagger \hat{a}_{\vec{k},l} \,,
\end{equation}
\ie $\braket{\varphi|\hat{N}|\varphi}$ is finite for each $\varphi \in [0]$. Therefore, this equivalence class alone represents the whole Fock space.

One can also understand the other equivalence classes in terms of particle number \cite{WS}. Two states are in the same equivalence classes if and only if their difference in particle number is finite:
\begin{equation}
   \ket{\varphi} \sim \ket{\Psi} \ \Leftrightarrow\ \braket{\varphi|\hat{N}|\varphi} - \braket{\Psi|\hat{N}|\Psi} < \infty \,,
\end{equation}
where it is understood that the subtraction is performed before the sum over the momentum modes. Since one can moreover show that each equivalence class is isomorphic to the Fock space, it follows that the von Neumann space can be thought of as infinite product of Fock spaces with unitarily inequivalent representations of the commutation relations in each subspace. So in each equivalence class $[\alpha]$, we have:
\begin{equation} \label{commutationRelationsVN}
	\left[\hat{a}_{l,\vec{k}}^{[\alpha]}, \hat{a}_{l',\vec{k'}}^{[\alpha]\dagger}\right] = \delta^{(3)}(\vec{k}-\vec{k}') \delta_{l l'} \,, \qquad \left[\hat{a}_{l,\vec{k}}^{[\alpha]}, \hat{a}_{l',\vec{k'}}^{[\alpha]}\right] = \left[\hat{a}_{l,\vec{k}}^{[\alpha]\dagger}, \hat{a}_{l',\vec{k'}}^{[\alpha]\dagger}\right] =0 \,.
\end{equation}
	This immediately raises the question what subspace of $\mathcal{H}_{\text{VN}}$ is physically relevant. A reasonable requirement for any state to be physical is that it contains finite energy. Whenever a theory has a mass gap, the Fock space -- defined by the requirement of finite particle number -- is the only equivalence class with finite energy and therefore contains all physically reasonable states. So it makes sense to restrict oneself to the Fock space. 

However, the situation is drastically different in a gapless theory. Then there can be states that contain an infinite amount of zero modes but nevertheless carry finite energy. Therefore, there are distinct equivalence classes with finite energy and there is no reason to restrict oneself to only one of them.
In a gapless theory, states of different equivalence classes are therefore physically sensible. In fact, as already noticed in \cite{WS} and emphasized recently in \cite{stromingerRevisited}, the $S$-matrix generically enforces the transition between different equivalence classes so that it is impossible to restrict oneself to a single equivalence class in an interacting system.  We will elaborate shortly on how this comes about. The fact that states in different equivalence classes are -- by definition -- orthogonal will be crucial for our discussion of IR-physics.

\subsection{Well-Defined Asymptotic States}
 In gapless theories such as QED and perturbative gravity, well-defined asymptotic states automatically contain an infinite number of soft photons/gravitons.
 The reason is that in the presence of long-range forces, asymptotic dynamics cannot be approximated by the free Hamiltonian, but the leading order of the interaction term also has to be taken into account \cite{FK}:\footnote
{We omit the Coulomb phase both in the definition of the asymptotic state and in the $S$-matrix since it will not matter for our discussion.}
\begin{equation}
	\W{t}_\lambda^r = \exp\left\{\frac{1}{\sqrt{2(2\pi)^3}} \limitint_\lambda^r \frac{\ldiff[3]{\vec{k}}}{ \sqrt{|\vec{k}|} }   \sum\limits_l \int \ldiff[3]{\vec{p}} \hat{\rho}(\vec{p})   \left(\frac{p\cdot \varepsilon^\star_{l,\vec{k}}}{p\cdot k} \hat{a}^\dagger_{l,\vec{k}} \ex^{i \frac{pk}{p_0}t} - \text{h.c.}\right) \right\}\,,
	\label{asymptoticDynamics}
\end{equation} 
where $\varepsilon^\mu_{l,\vec{k}}$ is the photon polarization vector and $\hat{\rho}(\vec{p})= e\sum\limits_s \left(\hat{b}^\dagger_{s,\vec{p}}\hat{b}_{s,\vec{p}}-\hat{d}^\dagger_{s,\vec{p}}\hat{d}_{s,\vec{p}}\right)$ is the charge density operator for electrons and positrons.\footnote
{Here $\hat{b}^\dagger_{s,\vec{p}}$/$\hat{d}^\dagger_{s,\vec{p}}$ is the creation operator for an electron/positron of spin $s$ and momentum $\vec{p}$.}
The limits of integration -- often left out in the literature -- are crucial for our treatment. That is why we explicitly indicate them in $\W{t}_\lambda^r$. The lower limit is an IR-regulator $\lambda$. As long as we keep it finite, we can work in the Fock space and the operator \eqref{asymptoticDynamics} is well-defined there. In the end, however, $\lambda$ will go to zero and it will turn out that this forces us to work in the larger von Neumann space. Whereas $\lambda$ is a regulator, it is clear that $r$ has to be non-vanishing since otherwise the operator \eqref{asymptoticDynamics} is trivial. So it is a finite and non-zero physical scale. 
As already introduced in \Eq \eqref{dressingIntro}, we can define asymptotic states by applying $\W{t}_\lambda^r$ to a bare state $\ket{\alpha}$ of electrons and positrons \cite{FK}:
\begin{equation} \label{asymptoticStateDefintion}
\ketD{\alpha}:=\hat{W}(t_{\text{obs}})_\lambda^r\ket{\alpha}\,,
\end{equation}
where  $t_{\text{obs}}$ is a so far arbitrary reference time.  We will keep it finite for now, but follow \cite{chung,FK} and set it to zero for the computation. The reason we can do so is that the final result only depends on the divergent zero-mode part of the dressing state whereas the phase controlled by $t_{\text{obs}}$ only changes the finite part of non-zero modes.\footnote
	{Strictly speaking, one can even by more general and choose an arbitrary state in the equivalence class $[\alpha]$ \cite{FK}. But since only the zero-mode part of dressing matters, we can adapt the choice \eqref{asymptoticStateDefintion} of \cite{chung,FK}. We will further comment on this freedom in choosing a dressing state in section \ref{ssec:comments}.}

Definition \eqref{asymptoticStateDefintion} also depends on $r$ and is non-trivial only for $r$ non-zero. Although we will keep $r$ general in our computation, we shall briefly discuss its physical interpretation.
If one wants to interpret $\ketD{\alpha}$ as initial or final state of scattering, it is most natural to think of $t_{\text{obs}}$ as the timescale after which the state will be measured. Once $t_{\text{obs}}$ is fixed, $r$ is no longer independent. The reason is the fact, noted in \cite{FK}, that the phases wash out if $k t_{\text{obs}}$ is sufficiently big, \ie $\lim_{t \gg k^{-1}} \exp\left(i k p t/p_0\right)/pk \approx 0$. Therefore, all modes with $k>t_{\text{obs}}^{-1}$ effectively disappear and do not contribute to the asymptotic dynamics any more:
\begin{equation} \label{cutW}
\W{t_{\text{obs}}}_\lambda^r \approx \W{t_{\text{obs}}}_\lambda^{t_{\text{obs}}^{-1}} \,.
\end{equation} 
Thus, if we only want to consider the physical modes, we have to set
\begin{equation}
r = t_{\text{obs}}^{-1} \,,
\label{rChoice}
\end{equation} 
\ie we can identify $r$ with the timescale $t_{\text{obs}}$ after which the final state is measured.
We can also justify the choice \eqref{rChoice} from a more physical point of view. Namely it is crucial for the photons in the dressing state that they are decoupled. Since a photon of energy $r$ needs a timescale of $r^{-1}$ to interact, it only makes sense to consider $r < t_{\text{obs}}^{-1}$. 
 While these arguments are heuristic, we will present a more precise justification for the choice \eqref{rChoice} in section \ref{sec:densityMatrix} by comparing the density matrix that we derive in our combined formalism with the result of \cite{decoherence1, decoherence2} that was obtained in a framework of real time evolution. We note, however, that our combined formalism is not tied to the physical interpretation of $r$ but works for an arbitrary choice.

 Before we investigate the dressed states more closely, we want to mention that the $S$-matrix is not modified in the dressed formalism \cite{FK}. The reason is that in the limit of infinite time, relation \eqref{cutW} becomes
		\begin{equation} \label{identityW}
	\lim_{t \rightarrow \infty} 	\W{t} = \idop \,,
	\end{equation}
which follows from $\lim_{t \rightarrow \pm \infty} \exp\left(i k p t/p_0\right)/pk =  \pm i \pi \delta(kp)$. For this reason, asymptotic dynamics do not contribute to the $S$-matrix but only modify the asymptotic states.
Setting $t_{\text{obs}}=0$, we get the asymptotic state \eqref{asymptoticStateDefintion}:
\begin{equation} \label{dressedState}
\ketD{\alpha} = \ket{\alpha}\otimes \Dad\,,
\end{equation}
where again we explicitly indicated the limits of integration.
The dressing $\Dad$ is the well-known coherent state of soft photons \cite{chung,kibble,FK}:
\begin{equation}
\Dad = \exp\left\{-\frac{1}{2}B_\alpha \ln\frac{r}{\lambda} \right\} \exp\left\{\limitint_\lambda^r \frac{\ldiff[3]{\vec{k}}}{\sqrt{|\vec{k}|}} \sum\limits_l \mathcal{F}^{(l)}_\alpha(\vec{k)} \,\hat{a}_{l,\vec{k}}^\dagger\right\}\ket{0} \,,
\label{dressingDefinition}
\end{equation}
where
\begin{equation}
	\mathcal{F}_\alpha^{(l)}(\vec{k}) =  \sum\limits_{n\in\alpha}\frac{e_n }{\sqrt{2(2\pi)^3}}\frac{p_n\cdot \varepsilon^\star_{l,\vec{k}}}{p_n \cdot k} \,.
\end{equation}
 The sum runs over all charged particles in $\alpha$ and $e_n$ is the charge of the $n^\text{th}$ particle. The state is normalized, \ie $\limitint_\lambda^r \frac{\ldiff[3]{\vec{k}}}{|\vec{k}|} \sum\limits_l|\FCa{\vec{k}}|^{2}= B_\alpha \ln\frac{r}{\lambda}$. This leads to
\begin{equation}
	B_\alpha = \frac{1}{2(2\pi)^3}  \sum\limits_{n,m\in \alpha}  \int \ldiff[2]\Omega \frac{e_n e_m \, p_n\cdot p_m}{p_n \cdot \hat{k} \ p_m\cdot \hat{k}}\,,
\end{equation} 
where $\hat{k}$ denotes the normalized 4-momentum of the photon.
When we investigate the particle number of the dressing state,
\begin{equation}
	\Da \hat{N} \Dad = B_\alpha \ln\frac{r}{\lambda}  \,,
\end{equation}
it becomes evident that it contains an infinite number of zero-energy photons in the limit $\lambda \rightarrow 0$. Thus, although the states possess the finite energy $B_\alpha r$, they are not in the equivalence class $[0]$, \ie in the Fock space. Note that varying $r$ does not change the equivalence class but only alters the energy of the dressing state. So the equivalence class only depends on the zero-momentum part of $\FCa{\vec{k}}$. 

In order to investigate how many different equivalence classes we have, we compute the overlap of two different dressing states:\footnote
{We use that $\sum_l \mathcal{F}_\alpha^{(l)\star}(\vec{k})\mathcal{F}_\beta^{(l)}(\vec{k})$ is real.}
\begin{align}
\prescript{r}{\lambda}{\braket{D(\alpha)| D(\beta)}}_\lambda^r = \, &\ex^{-\frac{1}{2} \limitint_\lambda^r \frac{\ldiff[3]{\vec{k}}}{|\vec{k}|} \sum\limits_l |\FCa{\vec{k}}|^{2} + |\FCb{\vec{k}}|^{2}} \nonumber \\ &\cdot \bra{0} \ex^{\limitint_\lambda^r\frac{\ldiff[3]{\vec{k}}}{\sqrt{|\vec{k}|}} \sum\limits_l\mathcal{F}_\alpha^{(l)\star}(\vec{k}) \hat{a}_{l,\vec{k}}}  \ex^{\limitint_\lambda^r \frac{\ldiff[3]{\vec{k}}}{\sqrt{|\vec{k}|}} \sum\limits_l \mathcal{F}_\beta^{(l)}(\vec{k})\hat{a}_{l,\vec{k}}^\dagger} \ket{0} \nonumber\\
= \, &\ex^{-\frac{1}{2}\limitint_\lambda^r \frac{\ldiff[3]{\vec{k}}}{|\vec{k}|} \sum\limits_l \left|\F{\vec{k}}\right|^2 } \,, 
\label{overlapUnintegrated}
\end{align} 
where we introduced the notation\footnote
{In all computations, results will solely depend on the difference $\F{\vec{k}}$. Therefore, it is possible to describe the same physical process with dressings that are shifted by a common function, $\FCa{\vec{k}} \rightarrow \FCa{\vec{k}} + C(\vec{k})$ and  $\FCb{\vec{k}} \rightarrow \FCb{\vec{k}} +  C(\vec{k})$. Such modifications of the dressing states have recently been considered in \cite{stromingerRevisited,akhouryDressing}.}
\begin{equation} \label{definitionF}
\F{\vec{k}} = \FCb{\vec{k}}- \FCa{\vec{k}} =\sum_{n\in \alpha, \, \beta}   \frac{e_n \eta_n }{\sqrt{2 (2\pi)^{3} }} \frac{p_n \cdot \varepsilon_{l,\vec{k}}^\star}{p_n \cdot k} 
\end{equation}
and $\eta_n=+1$ or $-1$ for an outgoing or incoming charged particle. We note that $\F{\vec{k}}$ is the same quantity that arises in the soft photon theorem \eqref{softTheorem} as a straightforward consequence of Taylor expanding the propagators of nearly on-shell electrons.
The radial integral is straightforward, $\limitint_\lambda^r \frac{\ldiff[3]{\vec{k}}}{|\vec{k}|} \sum\limits_l |\F{\vec{k}}|^{2}= \B \ln\frac{r}{\lambda}$, and the angular part gives
\begin{equation} \label{definitionB}
\B = \frac{1}{2(2\pi)^3}\sum_{n,\,m \in \alpha, \, \beta}\int \ldiff[2]{\Omega} \frac{\eta_n \eta_m e_n e_m\, p_n\cdot p_m}{p_n \cdot \hat{k}\ p_m \cdot \hat{k}} \,.
\end{equation}
This is a kinematical factor since it only depends on the initial and final state of scattering. For completeness, we note that it gives
\begin{equation}\label{Bintegraded}
\B= -\frac{1}{8 \pi^2} \sum_{n, \, m\in \alpha, \, \beta} \eta_n \eta_m e_n e_m \beta_{nm}^{-1} \ln \left(\frac{1+\beta_{nm}}{1-\beta_{nm}}\right) \,,
\end{equation}
where $\beta_{nm}$ is the relative velocity:
\begin{equation}
\beta_{nm} = \left(1-\frac{m_n^2 m_m^2}{(p_n \cdot p_m)^2}\right)^{1/2} \,.
\end{equation}

Using these integrals, we conclude that the overlap \eqref{overlapUnintegrated} of dressing states yields
\begin{equation}
\prescript{r}{\lambda}{\braket{D(\alpha)| D(\beta)}}_\lambda^r = \left(\frac{\lambda}{r}\right)^{\B/2}\,.
\end{equation}
It follows from \eqref{Bintegraded} that $\B=0$ only if the currents in $\ket{\alpha}$ and $\ket{\beta}$ match at each angle \cite{semenoff}. If this is not the case, $\Dad$ and $\Dbd$ have overlap zero for $\lambda \rightarrow 0$ and therefore are in different equivalence classes. Thus, there is a different equivalence class for each charge distribution on the sphere. We can parametrize the equivalence classes as $[\alpha]$ in terms of the charged states $\ket{\alpha}$.

\subsection{Equivalence Classes as Radiative Vacua}

In gapless theories, we have seen that non-trivial asymptotic dynamics lead to dressing states \eqref{dressingDefinition}, which -- in the limit $\lambda \rightarrow 0$ -- no longer belong to the Fock space because of an infinite number of zero-energy photons. However, as explained in section \ref{ssec:introductionVN}, each equivalence class of the von Neumann space is isomorphic to the Fock space. In particular, there is a representation of the commutation relations \eqref{commutationRelationsVN} in each of them \cite{WS}. We can formally relate them to the Fock space operators \eqref{commutationRelationsFock} via: 
\begin{equation} \label{equivalenceRepresentation}
    \hat{a}^{[\alpha]}_{l,\vec{k}} = \hat{W}(0) \hat{a}_{l,\vec{k}} \hat{W}^\dagger(0) \,.
\end{equation}
For finite $\lambda$, this representation is unitarily equivalent whereas it is not for $\lambda \rightarrow 0$. From the perspective of the operators $ \hat{a}^{[\alpha]}_{l,\vec{k}}$, the corresponding dressing state is a vacuum: 
 \begin{equation}
 	 \hat{a}^{[\alpha]}_{l,\vec{k}} \ketD{\alpha} = 0 \,.
 \end{equation}
So $\hat{a}^{[\alpha]\dagger}_{l,\vec{k}}$ represent excitations on top of the vacuum of the equivalence class $[\alpha]$, \ie  $\hat{a}^{[\alpha]\dagger}_{l,\vec{k}}$ corresponds to radiation on top of the dressing state defined by $\ket{D(\alpha)}$.

For $|\vec{k}|>r$, we have:
 \begin{equation}
 	 \hat{a}^{[\alpha]}_{l,\vec{k}} =  \hat{a}_{l,\vec{k}} \,,
 \end{equation}
 \ie photons of energy above $r$ are insensitive to the dressing and can be treated as if they were defined in the Fock space. As it will turn out explicitly in the calculation, only those photons constitute physical radiation. In contrast, photons of smaller energy solely occur in the dressing states but do not exhibit dynamics on their own. This is in line with the well-known decoupling of soft photons \cite{decoupling}.
 We remark that this is moreover consistent with the identification \eqref{rChoice} made above, $r=t_{\text{obs}}^{-1}$. Namely we expect that on the timescale $t_{\text{obs}}$, the softest radiation photons that can be produced have energy $t_{\text{obs}}^{-1}$, so all photons of smaller energy are decoupled.\footnote
 {That $t_{\text{obs}}^{-1}$ should correspond to an effective IR-cutoff for physical radiation was also proposed in \cite{decoherence1, decoherence2}.}

For our argument, however, the precise identification of the scale $r$ is inessential. The only important point is that $r$ splits the Hilbert space of photons in two parts. Photons below $r$ are part of the dressing. It is symmetric, \ie initial and final states are analogously dressed. Moreover, the dressing of the initial state is only sensitive to the initial state, but not to the final states and likewise for the final state. Since the dressing states contain an infinite amount of photons, they are not in the Fock space, but can only be defined in the larger von Neumann space. In contrast, photons above $r$ are part of radiation. It is asymmetric since we can prepare an initial state without radiation, \ie radiation only occurs in the final state but not in the initial state. In turn, it will become clear that it is sensitive to both the initial and the final state. In particular, it depends on the difference of initial and final state, \ie on the transfer momentum. The radiation state contains a finite number of photons and is well-defined in the Fock space. Thus, physical radiation is completely independent of the problems arising due to an infinite number of photons.

Radiation is characterized by a second scale $\epsilon$, which we can identify with the detector resolution. It is crucial to note that the scales $r$ and $\epsilon$ are in general independent  since they contain different physical information. The energy $r$ describes the timescale  after which the state is observed. In contrast, the scale $\epsilon$ corresponds to  the resolution scale of the particular device used to measure the final state. As explained, the only requirement is that $r<\epsilon$. In fact, it will turn out that $r \ll \epsilon$ is needed for a well-defined separation of dressing and radiation. In this limit, the energy carried by the dressing states is negligible. So all energy is carried by the radiation state whereas the only significant contribution to the number of photons comes from the dressing. In total, we obtain the following hierarchy of scales:
 \begin{equation}
 \lambda < r < \epsilon < \Lambda \,,
 \label{scaleHierarchy}
 \end{equation}
 where $\Lambda$ is the energy scale of the whole process, \eg the center-of-mass energy. In the existing literature, the scales $\lambda$, $\epsilon$ and $\Lambda$ are well-known. However, there is no additional scale $r$. The reason is that -- as we will show -- all rates are independent of $r$. So the introduction of the scale $r$, which separates dressing from radiation, is unnecessary if one is solely interested in rates. In contrast, it will turn out that the final density matrix does depend on $r$.  The reason is that unlike the rate, the density matrix depends on the timescale  after which it is measured. Therefore, we have to  keep the scale $r$ to derive an IR-finite density matrix.

Introducing the new scale $r$ amounts to interpolating between the well-known dressed and inclusive formalisms. We can consider the two limiting cases. For $r=\epsilon$, there is no radiation but all photons are attributed to dressing. This leads to Chung's calculation \cite{chung}, but corresponds to the unsatisfactory situation that there is no soft emission and that the resolution scale $\epsilon$ appears in the dressing of the initial state. The opposite limiting case is to set $r=\lambda$. Then there is no dressing, in particular the initial state is bare, but the final state contains photons of arbitrarily low energies. This leads to the calculations by Yennie, Frautschi and Suura \cite{YFS} as well as Weinberg \cite{weinberg}.  However, this construction lacks well-defined asymptotic states. For these reasons, we will work in the combined formalism that realizes the general hierarchy \eqref{scaleHierarchy}. We will demonstrate that doing so leads to the well-known IR-finite rates, but additionally it will allow us to obtain a well-defined density matrix of the final state.

\section{Combined Formalism}
\label{sec:amplitude}
\subsection{Calculation of Final State}

We consider a generic process of scattering. In order to determine the final state, we only need two ingredients: a well-defined initial state and the $S$-matrix of QED. Having defined the initial state \eqref{dressedState}, it remains to apply the $S$-matrix to it. The first step it to insert an identity. As shown in appendix \ref{app:identitiy}, it can be split in three parts if we assume $\epsilon \gg r$. The first one consists of photons with energy below $r$. Those are contained in the dressing of the hard states. The second one corresponds to radiative soft photons, \ie a state in which each single photon has an energy greater than $r$ but smaller than $\epsilon$.  As explained, the definition of a radiation photon generically depends on the radiative vacuum on top of which it is defined. However, it follows from \eqref{equivalenceRepresentation} that this distinction is inessential for photons of energy greater than $r$ and we can treat them as if they were defined in the usual Fock space. Finally, the third part consists of all remaining modes, \ie electrons and possibly hard photons.  As already introduced in \Eq \eqref{SMatrixIntro}, we therefore obtain a final state that consists both of dressed charged states and of radiation:
\begin{align} 
\label{calculationStart}
	\hat{S} \ketD{\alpha} 
	&= \sum_{\beta} \left(\frac{\lambda}{\Lambda}\right)^{\B/2} \ketD{\beta} \otimes \sum\limits_n \frac{1}{n!} \left(  \prod_{i=1}^n  \limitint_r^\epsilon  \ldiff[3]{\vec{k}_i} \sum_{l_i} \SD[\gamma_n]  \hat{a}_{l_i,\vec{k}_i}^\dagger \right) \ket{0}   \,,
\end{align}
where $n$ sums over the number of soft photons and the factor $1/n!$ comes from the normalization of the photon state.  The matrix element $\SD[\gamma_n]$ is evaluated between dressed electron states and moreover contains radiation in the final state: $\SD[\gamma_n] = \left(\braD{\beta} \hat{a}_{l_1,\vec{k}_1} \ldots \hat{a}_{l_n,\vec{k}_n} \right) \hat{S} \ketD{\alpha}$.

Corrections due to soft loops are included in the final state \eqref{calculationStart} and lead to the factor $\left(\lambda/\Lambda\right)^{\B/2}$, for which the first-order contribution was already announced in \eqref{1loop}. As discussed, the exponent $\B$, defined in \Eq \eqref{definitionB}, is non-negative and zero only for trivial scattering processes, in which the currents in $\ket{\alpha}$ and $\ket{\beta}$ match at each angle. Therefore, loop corrections only vanish for the zero-measure set of forward scatterings. In contrast, they lead to a vanishing amplitude for all non-trivial processes once we take $\lambda \rightarrow 0$.

 Now we can use that the soft photon theorem \eqref{softTheorem} holds in an arbitrary process to obtain $\SD[\gamma_n] = \braD{\beta} \hat{S} \ketD{\alpha} \prod_{i=1}^n \mathcal{F}_{\alpha,\,\beta}^{(l_i)}(\vec{k}_i)/\sqrt{|\vec{k}_i|} $, where the soft factor $\mathcal{F}_{\alpha,\,\beta}^{(l_i)}(\vec{k}_i)$ is displayed in \Eq \eqref{definitionF}. Moreover, we follow Chung's computation to evaluate the contribution of the dressing photons \cite{chung}:
\begin{equation} 
\braD{\beta} \hat{S} \ketD{\alpha} = \left(\frac{r}{\lambda}\right)^{\B/2} \S \,.
\end{equation}
So we obtain
\begin{equation}
\hat{S} \ketD{\alpha} = \sum_{\beta} \left(\frac{r}{\Lambda}\right)^{\B/2} \S \ketD{\beta} \otimes \sum\limits_n \frac{1}{n!} \left(  \prod_{i=1}^n  \limitint_r^\epsilon  \frac{\ldiff[3]{\vec{k}_i}}{\sqrt{|\vec{k}_i|}} \sum_{l_i} \mathcal{F}_{\alpha,\,\beta}^{(l_i)}(\vec{k}_i) \hat{a}_{l_i,\vec{k}_i}^\dagger \right) \ket{0}  \,.
\label{finalStateRaw}
\end{equation}
We can resum this final photon state:
\begin{equation}
\hat{S} \ketD{\alpha} = \sum_{\beta} \left(\frac{\epsilon}{\Lambda}\right)^{\B/2}  \S \left(\ketD{\beta} \otimes \Dabd\right)  \,,
\label{finalStateCoherent}
\end{equation}
where
\begin{equation} \label{radiationState}
\Dabd = \left(\frac{r}{\epsilon}\right)^{\B/2} \ex^{\limitint_r^\epsilon \frac{\ldiff{\vec{k}}}{\sqrt{|\vec{k}|}} \sum\limits_l \mathcal{F}_{\alpha,\beta}^{(l)}(\vec{k}) \, \hat{a}_{l,\vec{k}}^\dagger} \ket{0}
\end{equation}
is a normalized coherent radiation state and we used the integral \eqref{definitionB} to compute the norm. 

Formula \eqref{finalStateCoherent} makes the physics of the process very transparent. Both in the initial and in the final state, charged particles are dressed, as is required for well-defined asymptotic states. The dressings consist of photons of energy below $r$ and only depend on their respective state. This means that the dressing $\Dad$ of the initial state only depends on $\ket{\alpha}$ and the dressing $\Dbd$ of the final state only depends on $\ket{\beta}$. On top of the dressing, the final state (but not the initial state) also contains radiation. The radiation $\Dabd$ is made up of photons of energy above $r$ and depends both on the initial and on the final state of the hard electrons, and in particular on the momentum transfer between them.

As explained in the introduction, the main difficulty that arises from IR-physics -- which also seemingly leads to full decoherence -- comes from the fact that the dressing states are no longer in the Fock space due to the infinite number of zero-energy photons. For this reason, those states can only be defined in the much larger von Neumann space, which is isomorphic to an infinite product of Fock spaces. 
In our approach, we manage to separate this difficulty from the physical radiation. Namely only the dressing states $\Dad$ and $\Dbd$ contain an infinite number of photons, but these state do not correspond to physical radiation. Instead, they are part of the definition of asymptotic states. On top of the radiative vacuum defined by $\Dbd$, the radiation state $\Dabd$ exists. Since it only contains a finite number of photons of energies above $r$, it can be treated as if they were part of the usual Fock space. Only the radiation is measurable and for $r\ll \epsilon$, only it carries a significant energy.

We can check that the amplitude \eqref{finalStateCoherent} indeed gives the correct rate. To this end, we need to sum over all possible soft radiation in the final state, \ie over all radiation states in which the sum of all photon energies is below $\epsilon$. For $r\ll \epsilon$, we get
\begin{align}
\G &= \sum_n \frac{1}{n!} \left(\prod_{i=1}^n \limitint_r^\epsilon \ldiff[3]{\vec{k}_i} \sum\limits_{l_i}\right) \theta(\epsilon-\sum_{j=1}^n |\vec{k}_j|)  \left|\left(\bra{0}\hat{a}_{l_1,\vec{k_1}} \ldots \hat{a}_{l_n, \vec{k_n}}\otimes \braD{\beta}\right)\hat{S} \ketD{\alpha} \right|^2 \nonumber\\
& = \left(\frac{r}{\Lambda}\right)^{\B}   \sum_n \frac{1}{n!}\left(\prod_{i=1}^n \limitint_r^\epsilon \frac{\ldiff[3]{\vec{k}_i}}{|\vec{k}_i|} \sum\limits_{l_i} |\mathcal{F}_{\alpha,\,\beta}^{(l_i)}(\vec{k}_i) |^{2}\right) \theta(\epsilon-\sum_{j=1}^n |\vec{k}_j|) \left|\S\right|^2 \nonumber\\
& =  \left(\frac{\epsilon}{\Lambda}\right)^{\B} f(\B)  \left|\S\right|^2\,,\label{fullRate}
\end{align}
where energy conservation, encoded in the $\theta$-function, leads to \cite{YFS, weinberg}
\begin{equation} \label{definitionFFunction}
f(x) = \frac{\ex^{-\gamma x}}{\Gamma(1+x)} \,.
\end{equation}
Here $\gamma$ is Euler's constant and $\Gamma$ is the gamma function. This is the well-known result in the inclusive formalism \cite{YFS, weinberg}. If we neglect the function $f(\B)$, which is possible for weak coupling, the rate \eqref{fullRate} is also identical to the result in the dressed formalism \cite{chung}. In particular, it is clear that the answer that we obtain is IR-finite since the regulator $\lambda$ has dropped out. It is important to note that we never required IR-finiteness, but it simply arises as a consequence of applying the $S$-matrix to a well-defined initial state.

Moreover, we observe that the rate \eqref{fullRate} is also independent of the scale $r$. As we have discussed, our approach interpolates between the dressed formalism, which corresponds to $r=\epsilon$, and the inclusive formalism, which we obtain for $r=\lambda$.\footnote
{Sending $\epsilon\rightarrow r$ for fixed $r$ corresponds to a situation in which no soft emission takes place. When we work with well-defined, \ie dressed states, the rate of such a process is suppressed by the possibly small factor $\left(r/\Lambda\right)^{\B}$ but non-vanishing.}
The fact that our result is independent of $r$ implies that not only dressed and inclusive formalism yield -- except for $f(\B)$ -- the same rate, but that this is also true for our interpolation between them.

\subsection{Additional Comments}
\label{ssec:comments}
\subsubsection*{IR-Finite Amplitudes in Inclusive Formalism}
 Before we come to the study of the density matrix, we will briefly deviate from our main line of argument and make a few additional comments. First, we will for a moment take the limit $r=\lambda$, in which the dressings vanish and we obtain the inclusive formalism. Then formula \eqref{finalStateCoherent} becomes
\begin{equation}
\hat{S} \ket{\alpha} = \sum_{\beta} \left(\frac{\epsilon}{\Lambda}\right)^{\B/2}  \S \left(\ket{\beta} \otimes \ket{\gamma(\alpha, \beta)}_\lambda^\epsilon\right)  \,,
\label{finalStateInclusive}
\end{equation}
where the electron states are not dressed. This leads to the IR-finite amplitude:\footnote
{From this formula, the first-order contribution due to emission, which was already announced in \Eq \eqref{1emission}, is apparent.}
\begin{equation}
\left(\prescript{\epsilon}{\lambda}{\bra{\gamma(\alpha, \beta)}} \otimes \bra{\beta}\right)\hat{S} \ket{\alpha} = \left(\frac{\epsilon}{\Lambda}\right)^{\B/2}  \S \,.
\end{equation}
So if we use as final state the correct state of radiation $\ket{\gamma(\alpha, \beta)}_\lambda^\epsilon$, which depends both on initial and final electrons, we get an IR-finite amplitude in the inclusive formalism. However, the price we pay is that  on the one hand, we are not able to obtain the factor $f(\B)$ that encodes energy conservation and on the other hand that now the radiation state $\ket{\gamma(\alpha, \beta)}_\lambda^\epsilon$ contains an infinite number of zero-energy photons and is no longer part of the Fock space. Nevertheless, it is a physically sensible state since it only contains a finite energy.

\subsubsection*{Collinear Divergences}
An interesting question is what happens when one sends the mass of some of the hard particles to zero. In gravity, this situation is not special, \ie the kinematical factor $\B$ stays finite. This is connected to the fact that gravitational radiation is quadrupolar. In QED, however, the situation is drastically different. In the limit of a small electron mass $m$, it follows from \eqref{Bintegraded} that the exponent $\B$ scales as
\begin{equation} \label{collinearDivergence}
	\B  \sim -\ln m \,,
\end{equation}
\ie it becomes infinite for massless electrons. The only exception are processes of trivial scattering, in which the currents of initial and final state match antipodally. Thus, it is clear from  \eqref{fullRate} that the rate of any non-trivial scattering process vanishes.

 As a consequence, one could try to consider a wider class of processes such that a non-vanishing total rate can be obtained.  This was achieved in \cite{kln}, where -- on top of all soft emission processes -- a special class of emission {\it and absorption} processes was considered, namely the emission and absorption of {\it collinear} photons of arbitrary energy. Taking into account both emission and absorption is moreover required for Lorentz covariance: Unlike in the soft theorem \eqref{softTheorem}, a single amplitude of a particular emission process violates Lorentz invariance in the collinear limit. However, the physical soundness of adding absorption processes is questionable since the origin of the absorbed photons of arbitrary energy is unclear.

\subsubsection*{Large Gauge Transformations and Dressing}
Another interesting question is how gauge transformations act on dressed states. We can parametrize them as shift of the polarization tensor,\footnote
{In a pure $S$-matrix formalism, invariance under the shift \eqref{gaugeShift} can equivalently be derived from Lorentz invariance \cite{weinberg64}. Then charge conservation follows from the soft theorem \cite{weinberg64}: Plugging the shift \eqref{gaugeShift} in the soft factor \eqref{definitionF}, we conclude that
	\begin{equation} 
	\lambda_l^\star(\vec{k}) \sum_{n\in \alpha, \, \beta}   e_n \eta_n =0\,,
	\end{equation}
	\ie that the total incoming charge must be equal to the total outgoing charge. We note that this argument is completely independent of IR-divergences that arise due to soft loops.}
\begin{equation} \label{gaugeShift}
\varepsilon_{l,\vec{k}}^\mu \rightarrow \varepsilon_{l,\vec{k}}^\mu + \lambda_l(\vec{k}) k^{\mu} \,,
\end{equation}
where $\lambda_l(\vec{k})$ is an arbitrary function. Since dressing is determined by photons with $|\vec{k}|<r$,  small gauge transformations, for which $\lambda_l(\vec{k})$ vanishes for all $|\vec{k}|<r$, leave the dressing state invariant.
Only large gauge transformations, for which $\lambda_l(\vec{k})$ has support below $r$, act non-trivially. With the definition $\tilde{\lambda}_l(\vec{k}) = \lambda_l(\vec{k})\sum_{n\in \alpha} e_n /\sqrt{2(2\pi)^3}$, those lead to the transformed dressed state
\begin{align} \label{gaugeTransformationDressing}
\ketD{\tilde{\alpha}} = & \exp\left\{-\frac{1}{2}\limitint_\lambda^r \frac{\ldiff[3]{\vec{k}}}{|\vec{k}|}\sum_l \left|\mathcal{F}^{(l)}_\alpha(\vec{k}) + \tilde{\lambda}^\star_l(\vec{k}) \right|^2 \right\} \nonumber\\
& \exp\left\{\limitint_\lambda^r \frac{\ldiff[3]{\vec{k}}}{\sqrt{|\vec{k}|}} \sum\limits_l \left( \mathcal{F}^{(l)}_\alpha(\vec{k}) + \tilde{\lambda}^\star_l(\vec{k}) \right) \,\hat{a}_{l,\vec{k}}^\dagger\right\}\ket{\alpha} \,.
\end{align}
Thus, dressing states are not invariant under large gauge transformations\cite{kibble,FK}.\footnote
{In contrast, the radiation state \eqref{radiationState} is manifestly gauge-invariant.}

Since the number of photons only changes by a finite amount, the equivalence class to which the dressing state belongs and consequently also the cancellation of IR-divergences are left invariant. Instead, gauge transformations merely correspond to choosing a different representative of the equivalence class, \ie to modifying the choice \eqref{asymptoticStateDefintion}. However, the amplitude is not invariant under this transformation:
\begin{equation}
\braD{\beta}\hat{S} \ketD{\tilde{\alpha}} = \braD{\beta}\hat{S} \ketD{\alpha}\left(1-\frac{1}{2}\limitint_\lambda^r \frac{\ldiff[3]{\vec{k}}}{|\vec{k}|}\sum_l \left| \tilde{\lambda}_l(\vec{k}) \right|^2\right) \,,
\end{equation}
where we restricted ourselves to the leading order in $\tilde{\lambda}_l(\vec{k})$. This effect is weak for sufficiently small $ r\, \tilde{\lambda}_l(\vec{k})$ but generically non-zero. In order to restore full invariance, one has to apply the same shift \eqref{gaugeShift} to both initial and final states: $\braD{\tilde{\beta}}\hat{S} \ketD{\tilde{\alpha}} = \braD{\beta}\hat{S} \ketD{\alpha}$. This shows that dressing states do not exhibit dynamics on their own, but that -- in line with our previous discussion -- the physical meaning of dressing is to decouple photons of energy below $r$.

There is an interesting interpretation of the gauge transformed dressing state \eqref{gaugeTransformationDressing}. Up to an inessential phase factor, it can be written as
\begin{equation} 
\ketD{\tilde{\alpha}} \sim \exp\left\{-\frac{1}{2}\limitint_\lambda^r \frac{\ldiff[3]{\vec{k}}}{|\vec{k}|}\sum_l \left| \tilde{\lambda}_l(\vec{k}) \right|^2 \right\} \exp\left\{\limitint_\lambda^r \frac{\ldiff[3]{\vec{k}}}{\sqrt{|\vec{k}|}} \sum\limits_l   \tilde{\lambda}^\star_l(\vec{k})  \,\hat{a}^{[\alpha]\dagger}_{l,\vec{k}}\right\}\ketD{\alpha} \,,
\end{equation}
where $\hat{a}^{[\alpha]\dagger}_{l,\vec{k}}$ is defined in \Eq \eqref{equivalenceRepresentation}.
It becomes evident that large gauge transformations correspond to adding photons that are not defined in the Fock space, but according to the representation of the commutation relations in the equivalence class $[\alpha]$. The important point is that the representations of the large gauge transformations in each equivalence class are not unitarily equivalent. 
An interesting issue that goes beyond the scope of this paper is how in the case of gravity the generators of the Lorentz group are represented in the different equivalence classes and how this can be connected to BMS transformations. We hope to address this and other questions related to the interplay between dressing and large gauge transformations in a future work.

\section{Reduced Density Matrix}
\label{sec:densityMatrix}
\subsection{Well-Defined Tracing}
So far, we have rederived known results in a slightly different setting. Now we proceed to discuss the density matrix of the final state.  The crucial question is how to define the trace, \ie what states to trace over. But in our approach, in which we distinguish between dressing of asymptotic states and physical radiation, the answer is obvious. One should trace over soft radiation in the final state since it corresponds to physical states that are produced but not observed in a given setup. In contrast, the dressing is required for mere well-definedness of asymptotic states, so it makes no sense to trace over it. There is no asymptotic state without dressing.  Once the trace refers to physical radiation, it is well-defined in the Fock space because radiation only contains a finite number of photons.

Before tracing, the density matrix of the final state \eqref{finalStateCoherent} reads
\begin{align}
	\hat{\rho}^{\text{full}} & = \hat{S}  \ketD{\alpha} \braD{\alpha} \hat{S} \nonumber \\
	&= \sum_{\beta, \beta'} \left(\frac{\epsilon}{\Lambda}\right)^{\frac{\B+\Bp}{2}}  \S \Sp^* \left(\ketD{\beta} \otimes \Dabd \right) \left( \Dabp \otimes \braD{\beta'} \right) \,.
	\label{fullDensityMatrix}
\end{align}
Obviously, it is pure since no tracing has happened yet. Using an arbitrary basis $\{\ket{\gamma}\}_\gamma$ of radiation, \ie in the space of photons with energies above $r$ but below $\epsilon$, the trace is
\begin{align}
	\hat{\rho}^{\text{red}}  =\sum_{\gamma} \theta(E_\gamma - \epsilon) \sum_{\beta, \beta'} \left(\frac{\epsilon}{\Lambda}\right)^{\frac{\B+\Bp}{2}}  \S \Sp^* \ketD{\beta}\braD{\beta}  \braket{\gamma| \gamma(\alpha,\beta)}_r^\epsilon  \prescript{\epsilon}{r}{\braket{\gamma(\alpha,\beta')|\gamma}} \,,
	\end{align}
	where as in the computation of the rate, we imposed that the total energy $E_\gamma$ in radiation is at most $\epsilon$. If we neglect energy conservation for a moment, the computation becomes particularly transparent:
	\begin{align}
		\hat{\rho}^{\text{red}} \approx \sum_{\beta, \beta'} \left(\frac{\epsilon}{\Lambda}\right)^{\frac{\B+\Bp}{2}}  \S \Sp^*  \prescript{\epsilon}{r}{\braket{\gamma(\alpha,\beta')| \gamma(\alpha,\beta)}}_r^\epsilon\  \ketD{\beta}\braD{\beta}   \,.
\end{align}
Thus, we only have to compute the overlap of  coherent radiation states:
\begin{align}
 \prescript{\epsilon}{r}{\braket{\gamma(\alpha, \beta')| \gamma(\alpha, \beta)}}_r^\epsilon = \, & \ex^{-\frac{1}{2} \limitint_r^\epsilon \frac{\ldiff[3]{\vec{k}}}{|\vec{k}|} \sum\limits_l |\F{\vec{k}}|^{2} + |\Fp{\vec{k}}|^{2}} \nonumber \\ &\cdot \bra{0} \ex^{\limitint_r^\epsilon\frac{\ldiff[3]{\vec{k}}}{\sqrt{|\vec{k}|}} \sum\limits_l \mathcal{F}^{(l)*}_{\alpha,\beta'}(\vec{k}) \, \hat{a}_{l,\vec{k}}}  \ex^{\limitint_r^\epsilon \frac{\ldiff[3]{\vec{k}}}{\sqrt{|\vec{k}|}} \sum\limits_l \mathcal{F}_{\alpha,\beta}^{(l)}(\vec{k}) \, \hat{a}_{l,\vec{k}}^\dagger} \ket{0} \nonumber \\
=\, &\ex^{-\frac{1}{2}\limitint_r^\epsilon \frac{\ldiff[3]{\vec{k}}}{|\vec{k}|} \sum\limits_l \left|\F{\vec{k}}-\Fp{\vec{k}}\right|^2 } \nonumber\\
= \, &\ex^{-\frac{1}{2}\limitint_r^\epsilon \frac{\ldiff[3]{\vec{k}}}{|\vec{k}|} \sum\limits_l \left|\Fb{\vec{k}}\right|^2 }\nonumber \\
= \, & \left(\frac{r}{\epsilon}\right)^{\Bb/2} \,,
\end{align}
where the kinematical factor for a hypothetical process $\beta\rightarrow\beta'$ appeared.
In total, we obtain the element of the reduced density matrix:
\begin{align}
\rho^{\text{red}}_{\beta \beta'} &  = \left(\frac{\epsilon}{\Lambda}\right)^{\frac{\B+\Bp}{2}} \left(\frac{r}{\epsilon}\right)^{\frac{\Bb}{2}}  \S \Sp^*    \,,
\end{align}
where it is understood that indices refer to dressed states. Clearly, this result is IR-finite. Had we taken into account energy conservation, we would have gotten the result (which is a generalization of the computation in \cite{semenoff}):
\begin{align}\label{coherentDensityFull}
\rho^{\text{red}}_{\beta \beta'} &  = \left(\frac{\epsilon}{\Lambda}\right)^{\frac{\B+\Bp}{2}} \left(\frac{r}{\epsilon}\right)^{\frac{\Bb}{2}} f\left(\frac{\B+\Bp-\Bb}{2}\right)  \S \Sp^* \,.
\end{align}
As it should be, we observe that the diagonal terms reproduce the well-known rate \eqref{fullRate}, \ie $\rho^{\text{red}}_{\beta \beta} = \Gamma_{\alpha,\,\beta}$.

In previous work \cite{coherence1}, we already put forward an IR-finite density matrix using a more heuristic approach. We did so using an IR-finite version of the optical theorem and obtained a result similar to \eqref{coherentDensityFull}, but for $r=\epsilon$. Equipped with the arguments of the present paper, we can conclude that the prescription proposed in \cite{coherence1} corresponds to deriving the density matrix in the limit $r=\epsilon$, \ie in the dressed formalism, in which no decoherence occurs.\footnote
	{In \cite{coherence1}, we observed an additional factor proportional to $f(B)$, which we do not obtain in the present treatment. This additional factor lead to some decoherence. As we shall show shortly, however, this effect is subleading since the resulting decoherence is much smaller than the one we observe now for $r\ll\epsilon$.}

	 Finally, we can use the matrix element \eqref{coherentDensityFull} to justify our choice \eqref{rChoice} of $r$. In a framework of real time evolution, it was derived in \cite{decoherence1, decoherence2} for a particular simplified setup that the off-diagonal elements of the density matrix scale as $(1/t_{\text{obs}})^{\Bb/2}$, where $t_{\text{obs}}$ is the timescale after which the final state is measured. Comparing this with \eqref{coherentDensityFull}, we conclude that the identification $r\sim t_{\text{obs}}^{-1}$ was indeed justified. In this way, we obtain the same behavior as in \cite{decoherence1, decoherence2}: The longer we wait before we measure the final state, the smaller we have to choose $r$ and the more the off-diagonal elements of the density matrix get suppressed. We note, however, that our combined formalism does not rely on the identification \eqref{rChoice} and holds for general $r$.
	
	\subsection{Generalization to Superposition as Initial State}
	As suggested in \cite{semenoff3}, it is interesting to study a situation in which the initial state $\ket{\psi}$ is not a momentum eigenstate:
	 \begin{equation}
	\ketD{\psi} = \sum_\alpha f_\alpha^{(\psi)} \ketD{\alpha} \,,
	\end{equation}
	where $\sum_\alpha |f_\alpha^{(\psi)}|^2 = 1$ and we used the linearity of the definition \eqref{asymptoticStateDefintion} of dressing. Generalizing the above calculations, we get
	 \begin{align}\label{coherentDensitySuperposition}
	 \rho^{\text{red}}_{\beta, \beta'}
	 &=  \sum_{\alpha, \alpha'} f_\alpha^{(\psi)} f_{\alpha'}^{(\psi)*} \left(\frac{\epsilon}{\Lambda}\right)^{\frac{\B+ \Bpp}{2}} \left(\frac{r}{\epsilon}\right)^{\frac{\B+ \Bpp}{2}-B_{\alpha,\,\beta,\,\alpha',\,\beta'}} \nonumber\\
	 &\ \cdot f\left(B_{\alpha,\,\beta,\,\alpha',\,\beta'}\right)   \S S_{\alpha',\,\beta'}^* \,,
	 \end{align}
where
\begin{align}
	B_{\alpha,\,\beta,\,\alpha',\,\beta'} 
	&=  \frac{1}{2(2\pi)^3}\sum_{\substack{n\in \alpha, \, \beta\\ m\in \alpha', \, \beta'}}\int \ldiff[2]{\Omega} \frac{\eta_n \eta_m e_n e_m\, p_n\cdot p_m}{p_n \cdot \hat{k}\ p_m \cdot \hat{k}}\,.
\end{align}
The density matrix \eqref{coherentDensitySuperposition} applies to the most general case and thereby constitutes the main result of this section.

Clearly, this density matrix avoids full decoherence. In order to further analyze our result, we can decompose the sums:
 \begin{equation}
 	B_{\alpha,\,\beta,\,\alpha',\,\beta'} = \frac{\Bp + \Bpa -\Ba - \Bb}{2} \,.
 \end{equation}
This shows that if there is only one momentum eigenstate in the initial state, $f_\alpha^{(\psi)}=\delta_\alpha^{\alpha_0}$, the general density matrix \eqref{coherentDensitySuperposition} reduces to the result \eqref{coherentDensityFull} obtained before. It is moreover interesting to analyze the rates that we obtain:
\begin{align} \label{coherentDensitySuperpositionRate}
	\rho^{\text{red}}_{\beta, \beta}
	&=  \sum_{\alpha, \alpha'} f_\alpha^{(\psi)} f_{\alpha'}^{(\psi)*} \left(\frac{\epsilon}{\Lambda}\right)^{\frac{\B+ \Bpa}{2}} \left(\frac{r}{\epsilon}\right)^{\frac{\Ba}{2}} f\left(\frac{\B + \Bpa -\Ba }{2}\right)   \S S_{\alpha',\beta}^* \,.
\end{align}
	The $f(B)$-function is subleading since it follows from its definition \eqref{definitionFFunction} that it scales as $f(B)\sim 1-B^2$ for small $B$ and additionally it is insensitive to the ratio $r/\epsilon$. Therefore, we can focus on the other two IR-factors, $(\epsilon/\Lambda)^{(\B+ \Bpa)/2}$ and $(r/\epsilon)^{\Ba/2}$. Clearly, both are always smaller than $1$. In the case of constructive interference, they therefore always lead to a suppression of the rate. For destructive interference, however, they can work in both directions, \ie they can also serve to diminish suppressing contributions and thereby increase the rate. The $r$-dependent contribution $(r/\epsilon)^{\Ba/2}$ is particularly interesting since it does not factorize, \ie it cannot be absorbed in a redefinition of $\S$. These findings also hold for the off-diagonal elements. It is straightforward to show that the exponent of the $r$-dependent term in \Eq \eqref{coherentDensitySuperposition} is positive \cite{semenoff3}, so it also leads to a factor smaller than $1$. As before, this means that it leads to a suppression of off-diagonal elements if there is constructive interference. In particular, this is always the case when the initial state is only a single momentum eigenstate. In contrast, it can cause both suppression and enhancement for destructive interference.

	\subsection{Estimate of Amount of Decoherence}
 In order to quantify decoherence and the corresponding loss of information due to tracing over soft radiation, we compute the entanglement entropy of hard and soft modes. It is defined as the von Neumann entropy of the density matrix $\rho^{\text{red}}$ obtained after tracing over soft radiation. To estimate it, we apply the procedure developed \cite{coherence1}, where it was shown that a bound on the decoherence of $\rho^{\text{red}}$ can be given in terms of its distance to the nearest pure density matrix. To this end, we assume that we are given a pure density matrix $\rho^{\text{pure}} = \ketD{\Psi}\braD{\Psi}$ defined by a state 
	\begin{equation} \label{pureState}
		\ketD{\Psi} = \sum_{\beta }a_\beta  \ketD{\beta}\,,
	\end{equation}
which fulfills $|a_\beta|^2 = \rho^{\text{red}}_{\beta,\beta}$ but can have arbitrary phases. Thus, $\rho^{\text{pure}}$ and $\rho^{\text{red}}$ have the same diagonal and therefore describe the same rates. In this situation, it was derived in \cite{coherence1} that the entanglement entropy $S_{\text{soft}}$ is bounded by
	\begin{equation}\label{entropyBoundFormula}
		\frac{S_{\text{soft}}}{S_{\text{max}}} \lesssim \max_{\beta,\, \beta'}  |1-c^{(\Psi)}_{\beta,\, \beta'}|\,,
	\end{equation}
	where $S_{\text{max}}$ the maximal entropy that can exist in the Hilbert space and we defined 
		\begin{equation} \label{definitionC}
		c^{(\Psi)}_{\beta,\, \beta'}=	\frac{\rho^{\text{red}}_{\beta, \beta'}}{a_\beta a_{\beta'}^*} \,.
		\end{equation} 
		So the deviations of the $c^{(\Psi)}_{\beta,\, \beta'}$ from $1$ determine the decoherence and full coherence corresponds to $c^{(\Psi)}_{\beta,\, \beta'}=1$. 
	
 To derive a concrete bound on the entanglement entropy, we have to choose $a_\beta$. As said, the absolute value is fixed by the requirement $|a_\beta|^2 = \rho^{\text{red}}_{\beta,\beta}$. To obtain a bound that is maximally sharp, we therefore have to set the phases such that $c^{(\Psi)}_{\beta,\, \beta'}$ is minimal. In the explicit computation of $c^{(\Psi)}_{\beta,\, \beta'}$, it turns out that a good choice is\footnote
{The absolute value is fixed by rate \eqref{coherentDensitySuperpositionRate}. The phase is chosen such that it reproduces the density matrix \eqref{coherentDensitySuperposition} in the limit $r\rightarrow \epsilon$ and $f(B)\rightarrow 1$.}
	\begin{align}
a_\beta	=& \left(\sum_{\alpha, \alpha'} f_\alpha^{(\psi)} f_{\alpha'}^{(\psi)*} \left(\frac{\epsilon}{\Lambda}\right)^{\frac{\B+ \Bpa}{2}} \left(\frac{r}{\epsilon}\right)^{\frac{\Ba}{2}}  f\left(\frac{\B + \Bpa -\Ba}{2}\right)   \S S_{\alpha',\,\beta}^*\right)^{1/2}\nonumber\\
&\cdot \exp\left[i \arg \sum_{\alpha} f_\alpha^{(\psi)} \left(\frac{\epsilon}{\Lambda}\right)^{\frac{\B}{2}} \S\right] \,. 
	\end{align}
	Now we evaluate \eqref{entropyBoundFormula} in the regime of weak coupling where all kinematical factors become small, $B\ll1$. Then we can expand the exponential and the $f(B)$-functions. In the regime $r \ll \epsilon$, in which we work throughout, the contribution of the $f(B)$-functions is, as already explained, subleading and we will ignore it. Then we obtain to leading order
		\begin{align}
	\frac{S_{\text{soft}}}{S_{\text{max}}} \lesssim & \ln\frac{\epsilon}{r} \max_{\beta,\, \beta'} \frac{1}{2}\nonumber\\
	&\cdot \scalebox{1.5}{ \Bigg|} \frac{\sum_{\alpha, \alpha'} f_\alpha^{(\psi)} f_{\alpha'}^{(\psi)*} \left(\frac{\epsilon}{\Lambda}\right)^{\frac{\B+ \Bpp}{2}} \S S_{\alpha',\,\beta'}^*\left(\B+ \Bpp-2B_{\alpha,\,\beta,\,\alpha',\,\beta'}\right)}{\left( \sum_{\alpha} f_\alpha^{(\psi)} \left(\frac{\epsilon}{\Lambda}\right)^{\frac{\B}{2}} \S\right) \left( \sum_{\alpha} f_\alpha^{(\psi)*} \left(\frac{\epsilon}{\Lambda}\right)^{\frac{\Bp}{2}} \Sp^*\right)} \nonumber\\
	&- \frac{\sum_{\alpha, \alpha'} f_\alpha^{(\psi)} f_{\alpha'}^{(\psi)*} \left(\frac{\epsilon}{\Lambda}\right)^{\frac{\B+ \Bpa}{2}} \S S_{\alpha',\,\beta}^* \Ba}{2\left| \sum_{\alpha} f_\alpha^{(\psi)} \left(\frac{\epsilon}{\Lambda}\right)^{\frac{\B}{2}} \S\right|^2}\nonumber\\
	& - \frac{\sum_{\alpha, \alpha'} f_\alpha^{(\psi)} f_{\alpha'}^{(\psi)*} \left(\frac{\epsilon}{\Lambda}\right)^{\frac{\Bp+ \Bpp}{2}} \Sp S_{\alpha',\,\beta'}^* \Ba}{2\left| \sum_{\alpha} f_\alpha^{(\psi)} \left(\frac{\epsilon}{\Lambda}\right)^{\frac{\Bp}{2}} \Sp\right|^2} \scalebox{1.5}{\Bigg|}\,. \label{entropyBound}
	\end{align}
		Already at this point, the physical properties of this result become evident. First, decoherence depends logarithmically on the  ratio $\epsilon/r$. This means that it gets big if the resolution gets worse, \ie $\epsilon$ increases, or if one waits longer before measuring the final state, \ie $r$ decreases.\footnote
		{That the entropy due to tracing over soft modes should scale logarithmically with the resolution was already suggested in \cite{blackHole} using a simpler argument.}
		In the limit of the best achievable resolution, $\epsilon = r$, there is no decoherence.\footnote
		{Our derivation of the density matrix \eqref{coherentDensitySuperposition} relies on $r\ll \epsilon$ and is no longer valid in the limit $r=\epsilon$, which corresponds to the dressed formalism. However, it is easy to rederive the density matrix in this case and one obtains \eqref{coherentDensitySuperposition} but without the $r$-dependent factor and without the $f(B)$-function. Therefore, the bound \eqref{entropyBound} also holds for $r=\epsilon$ and shows that there is no decoherence in this limit.}
		 Moreover, we observe that the bound on the entanglement entropy scales with the kinematical factors $B$, \ie becomes small for small $B$-factors.\footnote
		 {An exception could occur in the case of fully destructive interference, \ie when one of the denominators in \Eq \eqref{entropyBound} vanishes. However, as long as only a small fraction of the entries of the density matrix goes to zero, it is clear that the amount of decoherence is still small. In that case, one would have to employ a more sophisticated bound than the one that we use here.}
		  Since they are proportional to $e^2$ and the momentum transfer, we conclude that decoherence scales with the coupling. Finally, it also depends on the kinematics of the scattering process. In the case in which the initial state is a single momentum eigenstates, this dependence becomes particularly transparent:
			\begin{equation}
			\frac{S_{\text{soft}}}{S_{\text{max}}} \lesssim\ln\frac{\epsilon}{r} \max_{\beta, \beta'}\frac{\Bb}{2} \,.
			\end{equation}
	Since the kinematical factor $\Bb$ depends on the angle between the electrons in $\beta$ and $\beta'$, we conclude that decoherence scales with the angle between different final states, \ie it gets bigger for bigger angles. This means that decoherence increases for final states whose bremsstrahlung is macroscopically different.

	\section{Summary and Outlook}
	\label{sec:conclusion}
	In this paper, we have given a first-principle derivation of the final scattering state in a theory that exhibits infrared divergences. Apart from the $S$-matrix, the only required input was a well-defined initial state. It contains the well-known dressing state of infinite photon number \cite{FK}, which does not belong to the Fock space but can only be defined in the larger von Neumann space \cite{VN}. Applying the $S$-matrix to this initial state, we obtained a final state that consists both of dressed charged states and of soft bremsstrahlung radiation. We performed the calculation in QED but the situation in gravity is fully analogous.
	
	The key point of our result is the distinction between dressing and radiation. The dressing is symmetric and independent of interaction, \ie initial and final states are analogously dressed and the dressing of the final state is independent of the initial state and vice versa. Therefore, dressing is by construction maximally entangled with the charged states. In contrast, radiation is asymmetric and sensitive to the interaction, \ie it only occurs in the final state and depends on the momentum transfer of the process. Consequently, it is not fully entangled with the final state.
	
	The distinction between dressing and radiation leads to an additional scale $r$. Using the results of \cite{decoherence1, decoherence2}, we can identify it with the inverse of the timescale after which the final state is measured. Photons of smaller energy  are decoupled and belong to the dressing whereas photons with higher energy, which are still below the resolution scale $\epsilon$, constitute soft radiation. For $r=0$, we recover the inclusive formalism \cite{BN,YFS,weinberg}, in which asymptotic states are not well-defined, and for $r=\epsilon$, we obtain the dressed formalism \cite{chung, kibble, FK}, which lacks soft emission. In this way, we interpolate between the two well-known formalisms. Since both yield the same rate, it is not surprising that our combined formalism does as well.

	The relevance of our approach lies in the fact that -- unlike the inclusive and dressed formalism -- it gives a well-defined meaning to the procedure of tracing over soft photons: One can only trace over unresolved soft radiation but not over dressing states, without which asymptotic states cannot be defined. This leads to an IR-finite density matrix with non-vanishing off-diagonal elements. It exhibits some decoherence due to tracing over unresolved radiation, but the amount of decoherence is generically small and therefore the density matrix is able to describe experimentally observed interference phenomena. In turn, the amount of decoherence observed in such experiments could be used  for a precise determination of the scale of dressing $r$.
	
	Finally, we want to mention that the different dressing states $\ket{D(\alpha)}$, which define {\it radiative vacua} in each von Neumann equivalence class $[\alpha]$, do not represent any vacuum degeneration in Goldstone sense. As stressed along the paper, transitions between these vacua are only possible with non-vanishing momentum transfer. Moreover, the equivalence classes are maximally sensitive to the charged states, \ie they do not contain more information than them and cannot be chosen independently.  Nevertheless, the different equivalence classes of the von Neumann space can acquire a meaning in the context of black holes. For a collapsing body, they could be interpreted as a bookkeeping tool to define quantum hair.

 \section*{Acknowledgements}
We thank Gia Dvali for interesting discussions. S.~Z.\ is grateful to the members of the High Energy Theory groups at EPFL Lausanne for many useful comments on \cite{blackHole, coherence1}. The work of C.G.\ was supported in part by Humboldt Foundation and by Grants: FPA 2009-07908 and ERC Advanced Grant 339169 "Selfcompletion". The work of R.L.\ was supported by the ERC Advanced Grant 339169 "Selfcompletion".

\appendix
\section{Split of Identity in Photon Sector}
\label{app:identitiy}
Our goal is to derive \Eq \eqref{calculationStart}. It originates from multiplying $\hat{S} \ketD{\alpha}$ with an identity that is decomposed as a tensor product of three factors. The first one, which we shall denote by $D$ and which will correspond to dressing, consists of all possible photons states composed of quanta with an energy below $r$. Analogously, the second one, which we shall call $\gamma$ and which will represent soft radiation, contains all possible photon states in which each photon has an energy above $r$ but below $\epsilon$. Finally, the third factor $\beta$ is composed of all remaining states, \ie photons with energy above $\epsilon$ and all other excitations, in particular charged particles. So we obtain:
\begin{equation}
		\hat{S} \ketD{\alpha} = \sum_{\substack{D\\( \lambda < E_D < r)}} \sum_{\substack{\gamma\\( r < E_\gamma < \epsilon)}} \sum_{\substack{\beta\\( \epsilon < E_\beta)}} \Big(\ket{\beta} \otimes\ket{\gamma} \otimes \ket{D}\Big)\Big(\bra{D} \otimes\bra{\gamma} \otimes \bra{\beta}\Big)\hat{S} \ketD{\alpha}\,.
\end{equation}

We will first turn to the sum over $D$. From Chung's computation \cite{chung} we know that $\Big(\bra{D(\beta)} \otimes\bra{\gamma} \otimes \bra{\beta}\Big)\hat{S} \ketD{\alpha} \neq 0$, \ie when we take the appropriate dressing $\ket{D(\beta)}$ of the final state $\ket{\beta}$, we obtain an IR-finite amplitude. (From the point of view of this computation, $\ket{\gamma}$ is a hard state.) This implies that any state $\ket{D}$ that belongs to a different equivalence class than $\ket{D(\beta)}$ has zero overlap with $\hat{S} \ketD{\alpha}$. In other words, the state in the mode $\vec{k}=0$, in which the number of photons is infinite, is fixed. In the identity, one would nevertheless have to perform independent sums over photons in the modes $0<|\vec{k}|<r$.\footnote
{In other words, as is discussed in \cite{FK}, one can replace $\FCa{\vec{k}}$ by $\FCa{\vec{k}} \varphi(\vec{k})$, where $\varphi(\vec{k})$ is an arbitrary function that fulfills $\varphi(\vec{k})=1$ in a neighborhood of $\vec{k}=0$. Then neglecting the sum over modes $0<|\vec{k}|<r$ corresponds to setting $\varphi(\vec{k})=1$ everywhere.}
However, if we take $r$ small enough, those mode do not change the result of $\hat{S} \ketD{\alpha}$ and we can  proceed as for \Eq \eqref{asymptoticStateDefintion} and fix them by the state $\ket{D(\beta)}$.  For $r\ll \epsilon$, we therefore obtain
\begin{equation}
	\sum_{\substack{D\\( \lambda < E_D < r)}} \ket{D}\bra{D} \approx \ket{D(\beta)} \bra{D(\beta)} \,.
\end{equation}
This means that the dressing is not independent but fixed by the hard state $\ket{\beta}$. In \cite{chung}, the same approximation is used, \ie the modes $0<|\vec{k}|<r$ are not treated as independent.

In contrast, we will not neglect any states in the sum over radiation. Writing it out explicitly, we get
\begin{equation}
\sum_{\substack{\gamma\\( r < E_\gamma < \epsilon)}} \ket{\gamma} \bra{\gamma} = 	\sum_{n} \frac{1}{n!}\left(\prod_{i=1}^n  \limitint_r^\epsilon  \ldiff[3]{\vec{k}_i} \sum_{l_i}\right) \Big(\hat{a}_{l_1,\vec{k}_1}^\dagger \ldots \hat{a}_{l_n,\vec{k}_n}^\dagger\ket{0}\Big) \Big(\bra{0}\hat{a}_{l_1,\vec{k}_1} \ldots \hat{a}_{l_n,\vec{k}_n}\Big) \,,
\end{equation}
where $1/n!$ comes from the normalization of the photon states. We will not resolve the third sum over hard modes $\beta$. In total, we obtain
\begin{equation}
		\hat{S} \ketD{\alpha} = \sum_{\beta}\sum_{n} \frac{1}{n!}\left(\prod_{i=1}^n  \limitint_r^\epsilon  \ldiff[3]{\vec{k}_i} \sum_{l_i}\right) \Big(\ketD{\beta} \otimes\ket{\gamma_n}\Big)\Big( \bra{\gamma_n} \otimes \braD{\beta}\Big)\hat{S} \ketD{\alpha} \,,
\end{equation}
where we introduced the notation $\ket{\gamma_n} = \hat{a}_{l_1,\vec{k}_1}^\dagger \ldots \hat{a}_{l_n,\vec{k}_n}^\dagger\ket{0}$.

\providecommand{\href}[2]{#2}\begingroup\raggedright\endgroup

\end{document}